\documentclass[aps,pra,twocolumn,superscriptaddress,amsmath,amssymb,floatfix]{revtex4-1}
\usepackage{graphicx}
\usepackage{bm}

\newcommand{\Op}[1]{\boldsymbol{\mathsf{\hat{#1}}}}
\def\openone{\leavevmode\hbox{\small1\kern-3.3pt\normalsize1}}

\begin{document}

\author{Marko Gacesa}
\email{gacesa@phys.uconn.edu}

\affiliation{Department of Physics, University of Connecticut, Storrs, CT 06269-3046, USA}
\affiliation{Max Planck Institute for the Physics of Complex Systems, 01187 Dresden, Germany}

\author{Subhas Ghosal}

\affiliation{Department of Chemistry, Birla Institute of Technology and Science Pilani, Hyderabad Campus, Jawahar Nagar, Shameerpet (M), Hyderabad, 500078, India}
\affiliation{Department of Physics, University of Connecticut, Storrs, CT 06269-3046, USA}

\author{Jason Byrd}

\affiliation{Department of Physics, University of Connecticut, Storrs, CT 06269-3046, USA}
\affiliation{Quantum Theory Project, University of Florida, Gainesville, Florida 32611}

\author{Robin C\^ot\'e}

\affiliation{Department of Physics, University of Connecticut, Storrs, CT 06269-3046, USA}
\affiliation{Institute for Quantum Computing, University of Waterloo, Waterloo, Ontario, Canada}

\title{Feshbach-optimized photoassociation of ultracold $^6$Li$^{87}$Rb molecules 
with short pulses}

\date{\today}

\begin{abstract} 
Two-color photoassociation of ground state $^6$Li$^{87}$Rb molecules via the $\mathrm{B}^1\Pi$ electronic state using short pulses near a magnetic Feshbach resonance is studied theoretically. A near-resonant magnetic field is applied to mix the hyperfine singlet and triplet components of the initial wave function and enhance the photoassociation rate, before the population is transferred to the ground state by a second pulse. We show that an increase of up to three orders of magnitude in the absolute number of molecules produced is attainable for deeply bound vibrational levels. This technique can be generalized to other molecules with accessible magnetic Feshbach resonances.
\end{abstract}

\maketitle

\section{Introduction}
Ultracold atomic and molecular gases offer unprecedented level of control over their internal and external degrees of freedom. As such, they present a multipurpose platform suitable for investigating fundamental phenomena in many-body physics inaccessible in other systems \cite{2010Sci...327..853O,2008PCCP...10.4079K,2002PhRvL..88f7901D,2006PhRvA..74e0301Y,2007PhRvL..98f0404B,2009RPPh...72h6401D,1367-2630-11-5-055049}.
Molecular gases, owing to their rich internal structure, are more versatile in this role than atomic gases, albeit at the cost of being more difficult to control. 

Ultracold diatomic molecules can be prepared in an ultracold atomic gas via photoassociation \cite{PhysRevLett.58.2420,RevModPhys.78.483,stwalley1999photoassociation} or magnetoassociation through a Feshbach resonance \cite{Kohler:RMP:2006,doi:10.1080/01442350600921772,2010RvMP...82.1225C}. 
This requires a sufficiently dense ultracold atomic gas, typically produced by laser cooling \cite{1996JChPh.104.9689B,metcalf1999laser} followed by evaporative cooling \cite{ketterle2002nobel}.
Neither of the two approaches allows direct production of large quantities of ultracold molecules in the rovibrational ground state. Magnetoassociation initially produces loosely bound ``Feshbach molecules'' close to the dissociation threshold and needs to be performed at very low temperatures in a tightly confined sample. Reaching such low temperatures is often a limiting factor and many experiments could benefit from a greater number of atoms and lower experimental complexity offered by magneto-optical traps (MOT). Magnetoassociation can be combined with optical coherent transfer, such as stimulated Raman adiabatic passage (STIRAP), to further cool down the molecules in one or more transitions towards the lowest vibrational levels \cite{2008PhRvA..78b1402K}. This approach was used successfully by Ni \textit{et al.} \cite{Ni10102008} to produce a sample of KRb molecules in the lowest vibrational levels. In subsequent experiments, using the same approach, Cs$_2$ molecules were produced in the lowest vibrational level \cite{danzl2009deeply}, whereas Rb$_2$ was produced successfully in the least energetic vibrational level of the lowest triplet electronic state \cite{lang2008ultracold}.

Photoassociation, being an all-optical method, is very general due to the large number of transitions available in diatomic molecules. Moreover, dynamic manipulation of laser fields allows enhanced control during the process \cite{koch2012coherent}. A major limitation of the technique, however, comes from the short lifetime of the excited electronic states and their poor relaxation rates into the lowest vibrational level of the ground state \cite{metcalf1999laser}. 
One possible approach to counter these limitations is to use short laser pulses, whose duration has to be significantly shorter than the spontaneous decay time of the excited state. 
In practice, this translates into laser pulses lasting tens of picoseconds or less. The first theoretical studies to explore photoassociation of molecules with short pulses \cite{2001PhRvA..63a3412V,2004EPJD...31..239L,2004PhRvA..70c3414L} focused on using a chirped pulse to control the population transfer to an excited state and dynamics of the wave packet to enhance the efficiency of radiative decay to the ground state. Koch \textit{et al.} \cite{PhysRevA.73.033408} considered a further optimization of the process by adding a second `dump' laser to transfer the population to the ground state. By optimizing the pulse parameters and time delay between the pulses, higher efficiency and a greater control over the population distribution in the ground state can be achieved. A similar approach, modified to exploit favorable spin-orbit coupling in the excited states, was considered by Ghosal \textit{et al.} \cite{SG:NJP:2009} to predict the formation efficiency of heteronuclear RbCs molecules. 

Another approach, investigated theoretically \cite{FOPA:PRL:2008} and experimentally \cite{PhysRevLett.101.060406,PhysRevA.87.052505}, relies on tuning an external magnetic field near a Feshbach resonance to alter the nodal structure of the free scattering wave function of the photoassociating atoms. The near-resonant wave function consists of strongly coupled bound and free hyperfine components of both singlet and triplet symmetry, resulting in greatly increased photoassociation rates with respect to selected vibrational levels. A similar mechanism was shown to increase the STIRAP efficiency to the point where a direct transition from the continuum to the ground state becomes  possible for experimentally attainable laser intensities \cite{Elena:NJP:2009}.

In this article, in an effort to increase the efficiency of production of ultracold molecules in the ground electronic state, we combine these two methods. We investigate photoassociation of ultracold $^6$Li$^{87}$Rb molecules in their ground state, performed by a pump-dump sequence of short laser pulses in a magnetic field tuned near a Feshbach resonance. This particular molecule has a comparatively strong electric dipole moment in its low-lying vibrational states of the electronic ground state \cite{Dulieu:JCP:2005}, making it a desirable candidate to manipulate with external electric fields. In addition, ultracold Li-Rb gas has been suggested as a suitable candidate for studying gas-crystal quantum transition \cite{Petrov:PRL:2007}, where heavy Rb atoms are thought to interact via an exchange of light Li atoms.
Ultracold Li-Rb mixtures were produced experimentally and Feshbach resonances at low and intermediate magnetic fields were detected and theoretically analyzed \cite{PhysRevLett.95.170408,Deh:PRA:2007,PhysRevA.78.022710,PhysRevA.79.012717,PhysRevA.79.042711,Ladouceur:09,PhysRevA.82.020701}. Increased interest in this system is also reflected in recently performed spectroscopic studies \cite{2011JChPh.134b4321I,Dutta2011,ivanova:094315,dutta2013}. Thus, Li-Rb represents a suitable system for this study.

The article is organized as follows. In Section \ref{sec:resonances} we perform a coupled-channel calculation to characterize magnetic Feshbach resonances in collisions of Li and Rb, as well as find a suitable resonance to use in two-photon photoassociation. In Section \ref{sec:tdwp} we discuss the proposed photoassociation scheme with respect to the electronic structure of LiRb molecule and numerically solve the time-dependent Schr\"odinger equation for a pump-dump laser pulse sequence. We optimize the pulses and determine the states' populations and total number of molecules formed. The results of \textit{ab-initio} calculation of electronic dipole transition moments used in this work are also given in this section. In Section \ref{sec:conclusion} we conclude and suggest possible directions for future research.

\section{Feshbach resonances and the initial state wave function}
\label{sec:resonances}

The photoassociation rate of molecules has been shown to be significantly enhanced near a Feshbach resonance \cite{FOPA:PRL:2008}, where the strong coupling of the open and closed hyperfine channels with respect to energy alters the nodal structure of the continuum wave function, resulting in the large increase of its amplitude in the short-range region. Consequently, its Franck-Condon overlap with the lower vibrational levels of electronic states is enhanced, giving rise to increased molecular formation rates. We rely on the same physical mechanism to enhance the population transfer to an intermediate excited electronic state during the pump step in two-photon photoassociation of LiRb.

The full multi-component continuum wave function is obtained by performing a quantum coupled-channel scattering calculation for $^6$Li and $^{87}$Rb atoms in an external magnetic field \cite{PhysRevB.38.4688}. 
We assume that the collisions take place at temperatures slightly above the quantum degeneracy conditions and limit our analysis to the $s$ and $p$ partial waves. 
We follow the procedure outlined below (see also Refs. \cite{PhysRevB.38.4688,PhysRevA.61.022721,Kohler:RMP:2006,2010RvMP...82.1225C}).

Two alkali atoms colliding in the magnetic field $\mathbf{B}=B \hat{\mathbf{z}}$ are accurately described by the two-body Hamiltonian \cite{1995PhRvA..51.4852M,PhysRevA.46.R1167} 
\begin{equation}
  \label{eq:H}
  \hat H = -\frac{1}{2 \mu R^2} \frac{\partial}{\partial R^2} R + \frac{\hat{\ell}^2}{2 \mu R^2} + 
      \sum_{j=1}^{2} \hat{H}^{\rm int}_j + \hat{V}_c \;,
\end{equation}
where $\mu$ is the reduced mass, $\hat{\ell}$ is the angular momentum operator, $\hat{H}^{\rm int}_j$ is the internal energy of atom $j$, and $\hat{V}_c$ is the electrostatic interatomic interaction. Additional higher-order terms, such as the magnetic dipole-dipole interaction, are neglected in this study. The atomic units are assumed throughout, if not explicitly written ($\hbar=1$). 

The internal Hamiltonian $\hat{H}^{\rm int}_j$ consists of the hyperfine and Zeeman interaction terms
\begin{equation}
  \hat {H}^{\rm int}_j= \alpha^{(j)}_{\rm hf} \mathbf{\hat{S}}_j \cdot \mathbf{\hat{I}}_j + 
  (\gamma_{e} \hat{S}_z^{(j)} - \gamma_n^{(j)} \hat{I}_z^{(j)}) B \; ,
\end{equation}
where $\mathbf{\hat{S}}_j$ and $\mathbf{\hat{I}}_j$ are the electronic and nuclear spin operators, respectively, $\alpha^{(j)}_{\rm hf}$ is the hyperfine coupling constant ($\alpha^{(1)}_{\rm hf} = 152.173$ MHz for $^6$Li and $\alpha^{(2)}_{\rm hf} = 3417.341$ MHz for $^{87}$Rb), $\gamma_e$ is the nuclear gyromagnetic ratio of the electron, and $\gamma_n^{(j)}$ is the nuclear gyromagnetic ratio ($\gamma_n^{(1)} = 0.822047$ $\mu_N$ for $^6$Li and $\gamma_n^{(2)} = 1.35298$ $\mu_N$ for $^{87}$Rb, where $\mu_N$ is nuclear magneton \cite{2005ADNDT..90...75S,NIST_codata}).

Within the Born-Oppenheimer approximation, the interaction potential $\hat{V}_c$ depends only on the internuclear separation $R$. It can be represented by the sum of lowest energy singlet and triplet molecular electronic potentials as
\begin{equation}
  \hat{V}_c(R) = V_0(R) \hat{P}^{(0)} + V_1(R) \hat{P}^{(1)} \; ,
\end{equation}
where the $\hat{P}^{(S)} = \sum_{m_S}{|S m_S \rangle \langle S m_S |}$ is the projection operator onto the singlet ($S=0$) or triplet ($S=1$) electronic spin configuration of the molecule with the interatomic potential $V_S(R)$ corresponding to the $X^1 \Sigma^{+}$ or $a^3 \Sigma^{+}$ state, respectively.

Molecular potentials are constructed by smoothly joining the short-range \textit{ab-initio} potentials to the long-range form given by the expansion \cite{Hirschfelder_book,1965JETP...21..624S}
\begin{equation}
   V_{\rm LR} = -\frac{C_6}{R^6}-\frac{C_8}{R^8}-\frac{C_{10}}{R^{10}} \pm 
                  A_{\rm ex} R^{a} e^{-b R} \; .
\label{eq:Vlr}
\end{equation}
We use the \textit{ab-initio} potentials by Korek $\emph{et al.}$ \cite{Korek:CP:2000}, the dispersion coefficients $C_6$, $C_8$, and $C_{10}$ from Derevianko $\emph{et al.}$ \cite{2001PhRvA..63e2704D,2003JChPh.119..844P}, and the exchange energy parameters $A_{\rm ex}=0.0058$, $a=4.9417$, and $b=1.1836$. All parameters are given in atomic units. The connection to the long-range form is performed at 13.5 Bohr.

\begin{figure}[tb]
 \centering
 \includegraphics[clip,width=\linewidth]{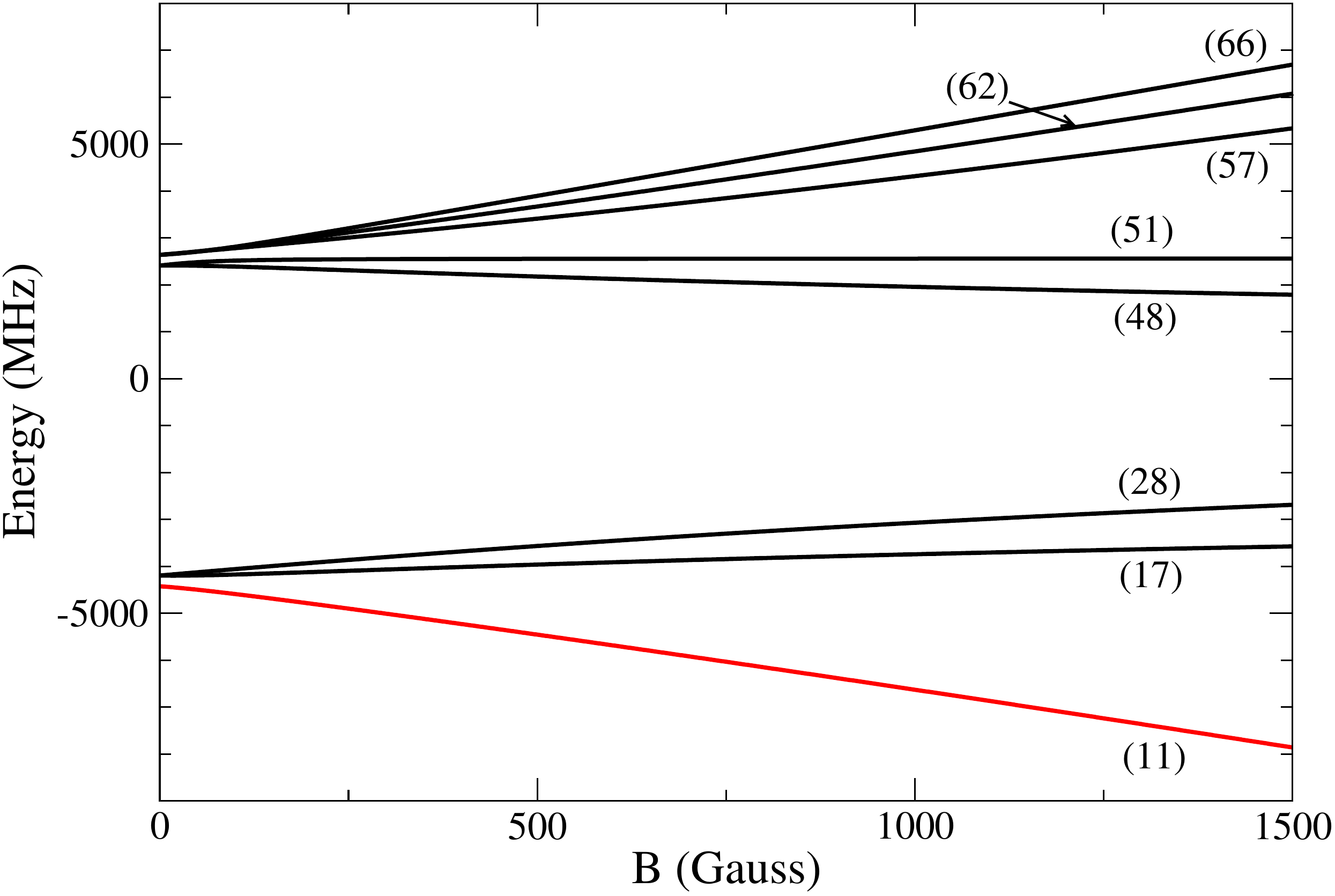}
 \caption{Zeeman splitting of the hyperfine levels coupled to the lowest-energy hyperfine state of the $^6$Li$^{87}$Rb molecule. The states are labeled in the channel basis $(\alpha \beta)$. The initial scattering state is shown in red.}
 \label{fig:hyperfine}
\end{figure}

The total wave function for the atoms prepared in the initial channel $\chi$ is expanded in the long-range uncoupled hyperfine basis set \cite{1965tac..book.....M,PhysRevB.38.4688,PhysRevA.61.022721,Kohler:RMP:2006,2010RvMP...82.1225C}
\begin{equation}
 |\psi\rangle = \frac{1}{R} \sum_{\chi} \varphi_{\chi}(R) |\chi\rangle ,
 \label{eq:psi_coupled}
\end{equation}
where $\chi=(F_1 \, m_{F_1} \, F_2 \, m_{F_2})$, $\varphi_{\chi}(R)$ are the corresponding expansion coefficients. Here, $\mathbf{F}_j$ is the total spin of the atom $j$ defined as $\mathbf{F}_j = \mathbf{S}_j + \mathbf{I}_j$ and $m_{F_j}$ is its projection on the internuclear axis.
In the separated atoms limit the basis can be represented as the direct product of atomic hyperfine states  
\begin{equation}
 |\chi \rangle = \vert F_1 \,m_{F_1} F_2 \,m_{F_2}\rangle \equiv \vert F_1 \, m_{F_1}\rangle_{\rm Li} \otimes \vert F_2 \, m_{F_2}\rangle_{\rm Rb} .
 \label{eq:chi}
\end{equation}

For small internuclear separations the molecular basis $\vert S I F m_F\rangle$ becomes more appropriate, where $\mathbf{S}=\mathbf{S}_1+\mathbf{S}_2$, $\mathbf{I}=\mathbf{I}_1+\mathbf{I}_2$, $\mathbf{F}=\mathbf{F}_1+\mathbf{F}_2$, and $m_F$ is its projection. 
To simplify the labeling of the hyperfine states we also use the ``channel basis'' $\vert \alpha \beta \rangle$, with the six hyperfine states of $^6$Li labeled as $\alpha = 1\ldots 6$ and eight hyperfine states of $^{87}$Rb as $\beta = 1\ldots 8$, in the order of increasing energy \cite{PhysRevA.61.022721,Kohler:RMP:2006}, such that \textit{e.g.} the channel $|11\rangle$ corresponds to the lowest energy state of both atoms
\begin{equation}
  |11\rangle = \vert \frac{1}{2}\, \frac{1}{2}\, 1\, 1 \rangle = | \frac{1}{2}\, \frac{1}{2} \rangle_{\rm Li} \otimes |1\, 1\rangle_{\rm Rb}.
\end{equation}
Note that the channels are sometimes labeled using lowercase roman letters.

As a consequence of the preserved rotational symmetry around the internuclear axis, the projection $m_F = m_{F_1} + m_{F_2}$ remains invariant throughout the collision \cite{PhysRevB.38.4688,PhysRevA.61.022721}, significantly reducing the number of coupled channels. We assume the scattering to take place with both atoms in their lowest hyperfine states, for which $m_F=3/2$, resulting in a total of eight coupled channels. 
To obtain a simple picture of the channels' dependence on the magnetic field, we solve the asymptotic form of Eq. (\ref{eq:H}), where the internuclear potential $V(R)$ is neglected (Figure \ref{fig:hyperfine}).

\begin{figure}[tb]
 \centering
 \includegraphics[clip,width=\linewidth]{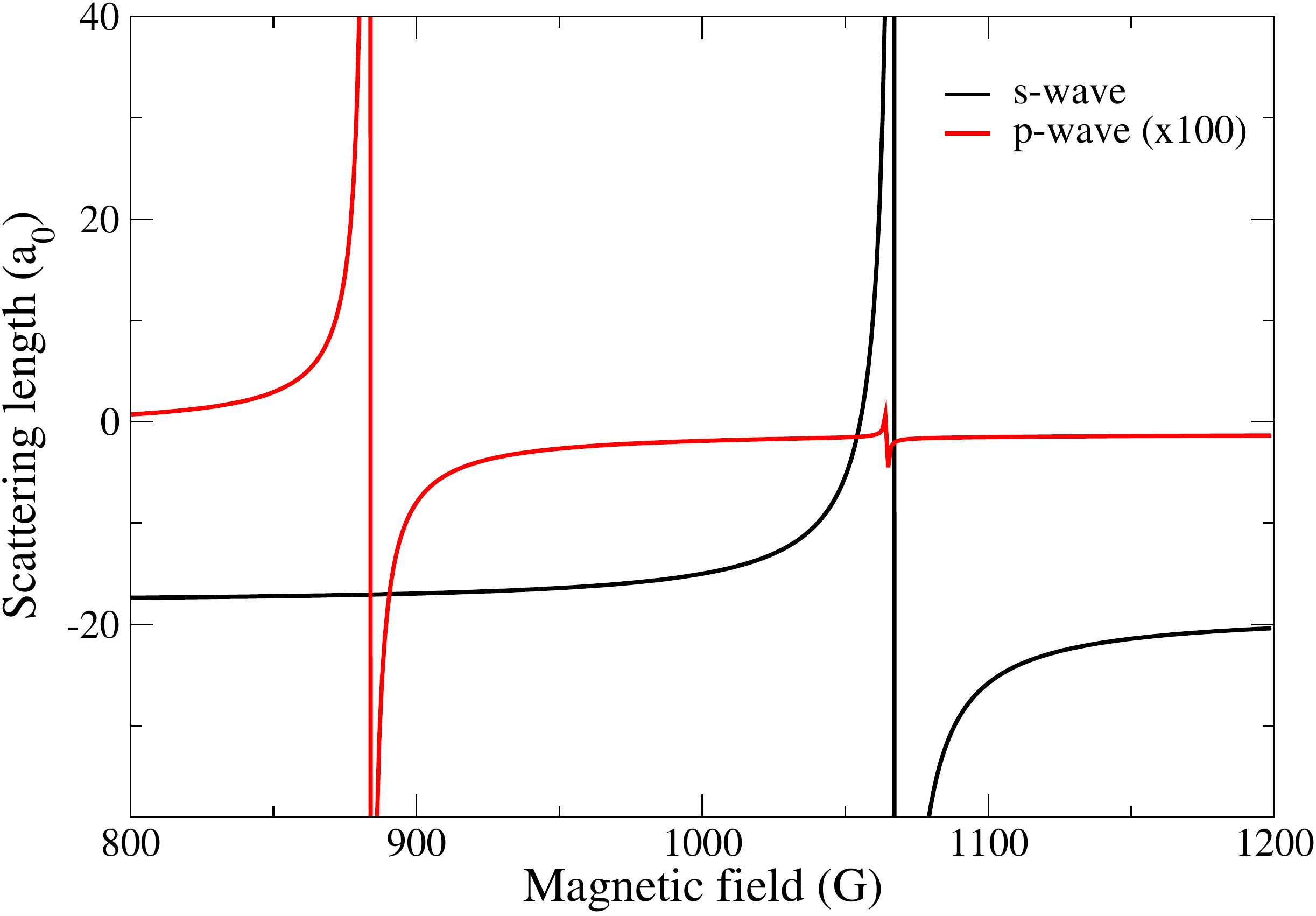}
 \caption{Scattering length as a function of the magnetic field for the least energetic hyperfine initial channel. The $s$-wave (black) and $p$-wave (red) are shown.}
 \label{f:resonances}
\end{figure}

Once the bases are defined, the full coupled-channel form of Eq. (\ref{eq:H}) is solved using a multichannel implementation of renormalized Numerov propagator method \cite{1977JChPh..67.4086J}.
The scattering length $a$ is obtained from the low-energy phase shift of the initial entrance channel $\eta_0(k)$ according to \cite{1965tac..book.....M,2010RvMP...82.1225C}
\begin{equation}
   \tan \eta_{0}(k) = -k a
\end{equation}
where $k = \sqrt{2\mu E}$ is the wave number associated with the pair of colliding atoms of relative energy $E$. This procedure is repeated for a range of magnetic fields until the satisfactory resolution is reached.

In order to reproduce the Feshbach resonances reported in a mixture of ultracold $^6$Li and $^{87}$Rb gases \cite{Deh:PRA:2007,PhysRevA.78.022710,PhysRevA.79.042711}, we constructed model potentials by adjusting the slope of the repulsive inner wall of the singlet and triplet potential curves by shifting the data points for the internuclear separations smaller than the equilibrium point, $R<R_{\rm eq}$, according to 
\begin{equation} 
  R_{{\rm shifted}} = R + x_{\rm S} \frac{R - R_{\rm eq}}{R_c-R_{\rm eq}} \; , 
\end{equation} 
where $R_c$ is the classical turning point and $x_S$ is the shift parameter for the singlet ($S=0$) and triplet ($S=1$) curve. The best agreement was reached for $x_0 = -0.103$ $a_0$ and $x_1 = 0.0315$ $a_0$. 

The calculated $s$-wave and $p$-wave Feshbach resonances for the atoms in the least energetic hyperfine channel $|11\rangle$ are shown in Figure \ref{f:resonances}. 
The resonance position $B_0$ and width $\Delta B$ are calculated by fitting the scattering length to the well-known form \cite{PhysRevA.51.4852}
\begin{equation}
   a(B)=a_{\rm bg} \left( 1 + \frac{\Delta B}{B-B_0} \right),
\end{equation}
resulting in the fit parameters $B_0 = 1067.85$ G, $\Delta B = 12.64$ G, and $a_{\rm bg}=-18.48 \,a_0$
for the broader $s$-wave resonance. 
In the $p$-wave scattering length we find two resonances at 884.6 G and 1065 G, confirming the prediction of Li \textit{et al.} \cite{PhysRevA.78.022710}. The first resonance has been detected experimentally \cite{Deh:PRA:2007} and later identified as a $p$-wave resonance \cite{PhysRevA.78.022710,PhysRevA.79.012717}, while the much narrower second $p$-wave resonance overlaps with the broad $s$-wave resonance (Figure \ref{f:resonances}). Our results are comparable to the ``Model II'' in Marzok \textit{et al.} \cite{PhysRevA.79.012717}, even though we make no additional efforts to achieve the level of accuracy required to discern between the proposed models.

\begin{figure}[tb]
 \centering
 \includegraphics[clip,width=8.5cm]{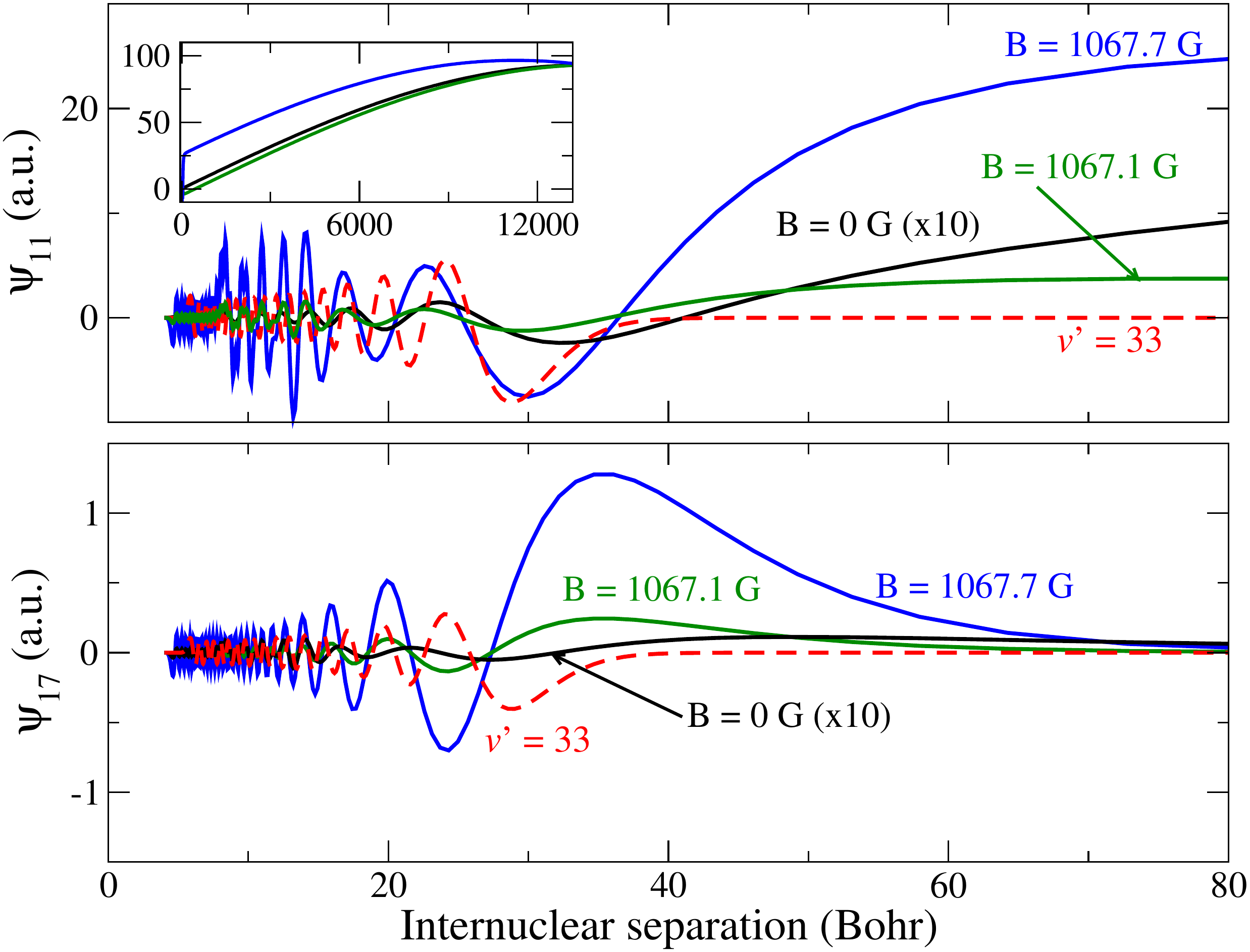}
 \caption{\textit{Top:} Open singlet component $|11\rangle$ of the scattering wave function for $B=0$ (black) and two near-resonant values of the magnetic field (green and blue curves). The wave function for the vibrational level $v'=33$ of the B$^1\Pi$ state is also shown (red). The wave functions are normalized to the relative momentum consistently between the components. \textit{Inset:} Long range part of the scattering wave functions.
 \textit{Bottom:} Closed singlet component $|17\rangle$ of the scattering wave function shown for the same values of magnetic field as above. 
 \label{fig:wfs_singlet}
}
\end{figure}

Using the adjusted molecular potentials we calculated the scattering wave functions as functions of the magnetic field for the initial scattering channel $|11\rangle$. The wave functions for selected values of the magnetic field are shown in Figure \ref{fig:wfs_singlet}. A large enhancement of the amplitude is clearly visible for all channels within a few tens of Gauss from the broad Feshbach resonance. 

The photoassociation rate of molecules depends on the dipole transition matrix element between the initial scattering state and the target bound vibrational level. Consequently, the enhanced wave function amplitude at short and medium internuclear distances can result in enhanced Franck-Condon overlap and molecular formation rate for suitable target levels. To illustrate this, we plot the wave function for the vibrational level $v'=33$ of the B$^1\Pi$ electronic state alongside the scattering wave function components in Figure \ref{fig:wfs_singlet}. Note the shift in the nodal structure between $B=0$ G and the near-resonant wave functions.

\section{Time-dependent wave packet calculations}
\label{sec:tdwp}

\subsection{Photoassociation scheme}

\begin{figure}[t]
\includegraphics[clip,width=\linewidth]{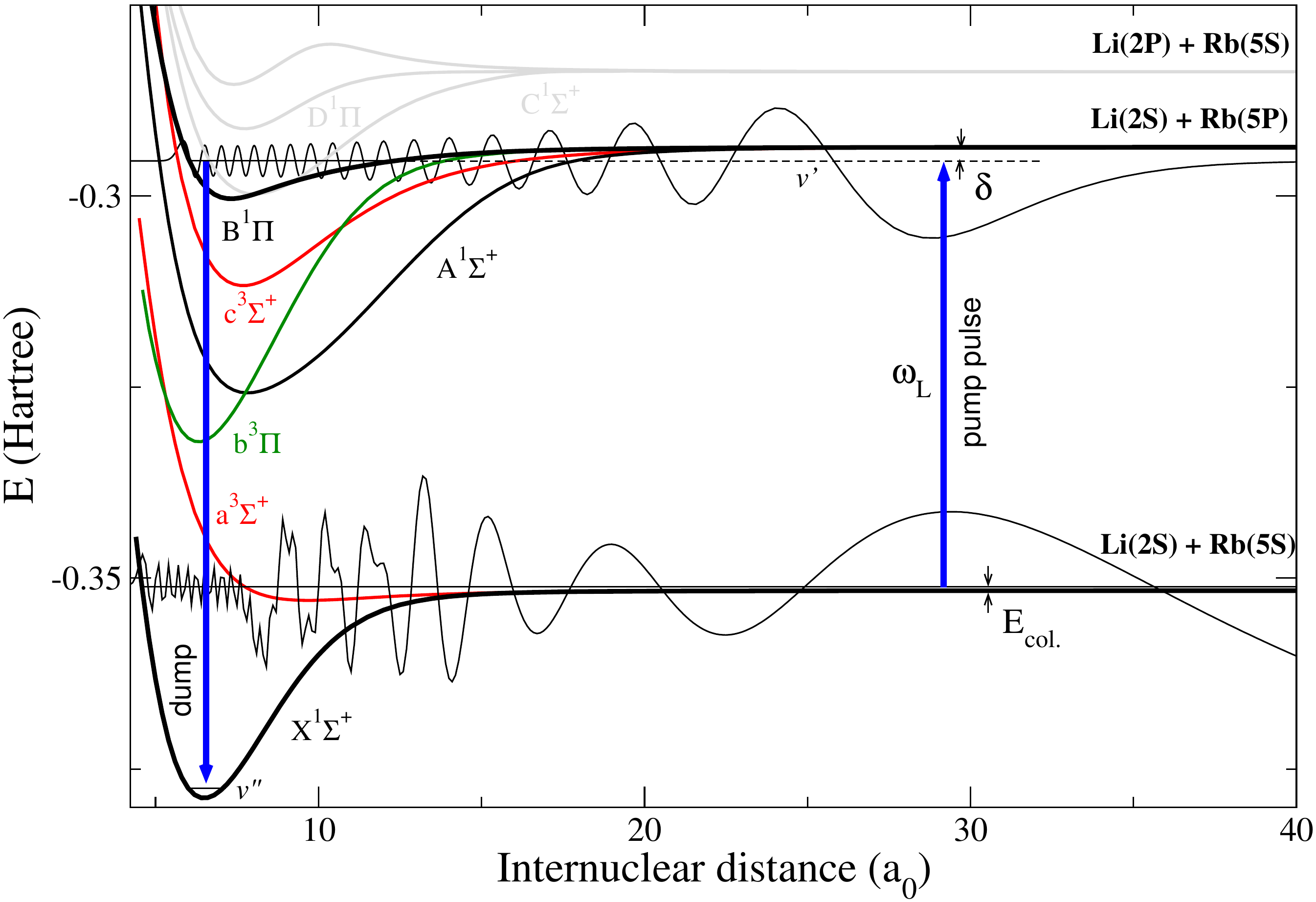}
\caption{Two-photon photoassociation process in LiRb. Potential energy curves and the wave functions are shown for reference. The red-detuned pump pulse photoassociates molecules in the excited $\mathrm{B}^1\Pi$ state, forming a wave packet that propagates inwards. When the wave packet reaches a favorable overlap with the deeply bound vibrational levels of the X$^1\Sigma^+$ state the second pulse is activated to transfer the population.
}
\label{f:scheme}
\end{figure}

The proposed two-photon photoassociation pump-dump sequence is illustrated in Figure \ref{f:scheme}. The pump pulse, red-detuned by $\delta$ from the Li(2$S$)+Rb(5$P$) asymptote, photoassociates LiRb molecules and forms an inward-propagating wave packet in the excited B$^1\Pi$ state. The spectral width of the pulse needs to be sufficiently large to excite several vibrational levels. 
After the optimal time delay, the second pulse is initiated to transfer the population into deeply bound vibrational levels of the ground X$^1\Sigma^+$ state, forming stable ultracold LiRb molecules.
The required time scale for both pulses is in the picosecond regime.

Our choice of B$^1 \Pi$ state as the intermediate state is based on recent spectroscopic studies of $^6$Li$^{87}$Rb molecules \cite{Dutta2011,2011JChPh.134b4321I,ivanova:094315}. The spectroscopic results agree well with earlier \textit{ab-initio} structure calculations \cite{Korek:CP:2000}, and both sets of constructed potential energy curves show favorable Franck-Condon overlap of the near-threshold vibrational levels of B$^1 \Pi$ electronic state with low-lying vibrational levels ($v'$ = 0 - 5) of the ground state. 
This is of critical importance for the proposed photoassociation scheme as the total population transfer to deeply bound vibrational levels strongly depends on the efficiency of the dump pulse.

% Photoassociation scheme

The spin-orbit coupling splits the Li(2$S$) + Rb(5$P$) asymptote into 2S$_{1/2}$ + 5P$_{1/2}$ and 2S$_{1/2}$ + 5P$_{3/2}$ in the long range (internuclear separation larger than the LeRoy radius) part of the potential. These asymptotes correspond to A$^1\Sigma^+$, b$^3\Pi$, B$^1\Pi$ and c$^3\Sigma^+$ electronic states in the Hund's case (a) coupling. For larger internuclear separations the spin-orbit coupling is significant and the B$^1 \Pi$ state becomes a mixture of singlet and triplet states.
However, in the separation range relevant to our pump-dump scheme (see Fig. 4), the spin-orbit coupling is negligible and the excited electronic state is well represented by the B$^1 \Pi$ state \cite{Korek:Theochem:2009,ivanova:094315}. 
Hence, to a good approximation, we consider the B$^1 \Pi$ state to be correlating with the
Li(2$S$ $^2S_{1/2}$) + Rb(5$P$ $^2P_{1/2}$) asymptote. 
We use the spin-orbit corrected interaction potential of Korek $\emph{et al.}$ \cite{Korek:Theochem:2009} for the excited B$^1\Pi$ electronic state. 

Note that the perturbation of the B$^1 \Pi$ state caused by the rotational coupling \cite{lefebvre2004spectra} with the C$^1 \Sigma^+$ state, correlating to the Li(2$s$) + Rb(4$d$) asymptote, is possibly significant for the low-lying vibrational levels ($v'= 2, 3$) of the B$^1 \Pi$ electronic state \cite{Dutta2011}, potentially enabling resonant interaction between these states. The rotational coupling is negligible for the higher vibrational levels and neglected in the present study.
%check if 4d is correct!

\subsection{Electronic dipole transition moments}
% by Jason Byrd

\begin{figure}[b]
 \centering
 \includegraphics[clip,width=\linewidth]{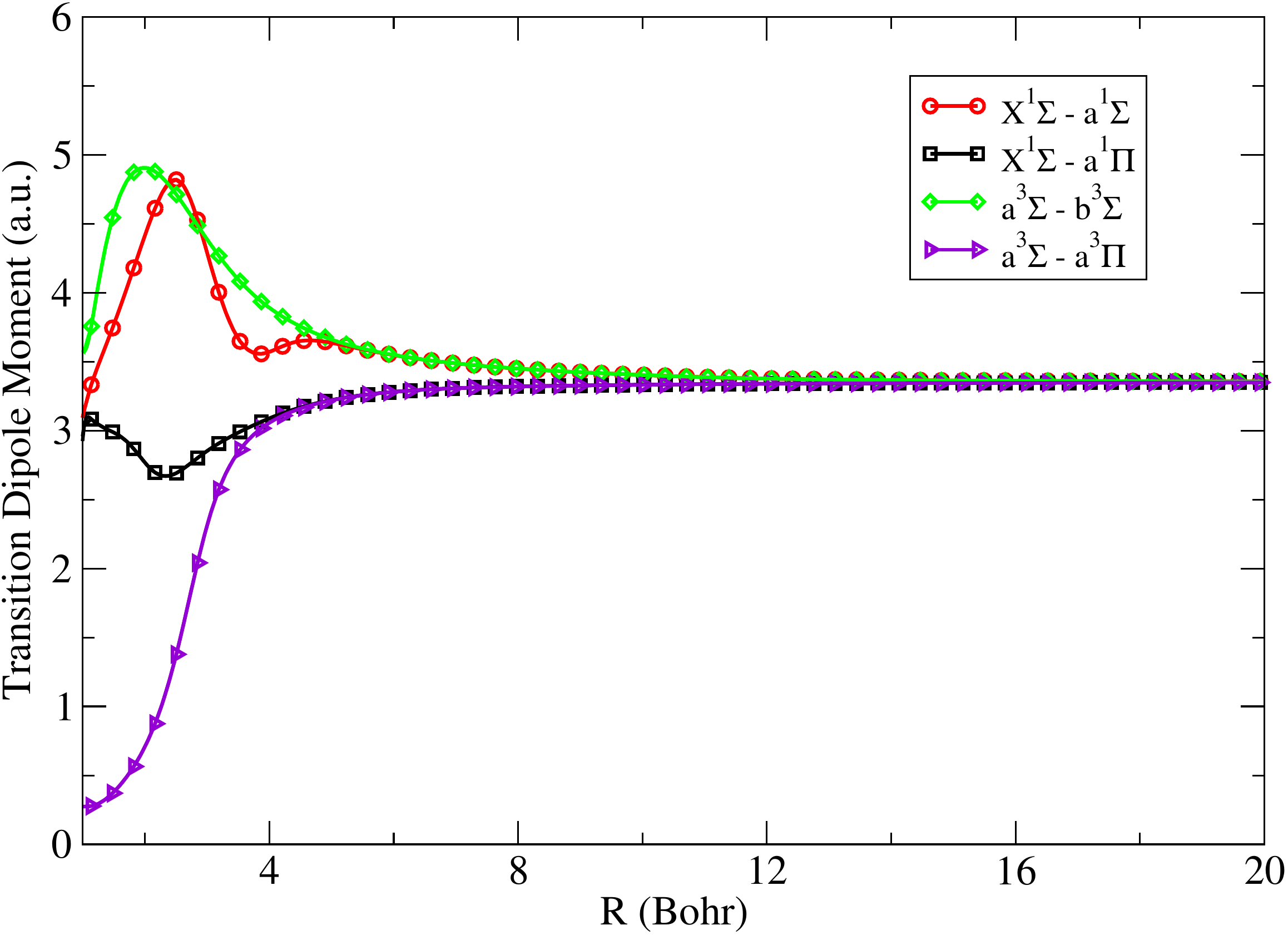}
 \caption{Electronic dipole transition moments of LiRb constructed in this work.}
 \label{fig:dipole}
\end{figure}
To compute the transition probability between initial, intermediate, and final states, electronic dipole transition moments are required. Due to the lack of available published data, we computed the electronic dipole transition moment between lowest singlet and triplet electronic states of LiRb and singlet and triplet electronic states converging to the Li$(2S)$+Rb$(5P)$ asymptote.

The electronic dipole transition moments as a function of internuclear separation (shown in Figure~\ref{fig:dipole}) were calculated at the full configuration interaction level of theory using the internally contracted multireference configuration interaction with single and double excitations \cite{werner1988,knowles1988} program implemented in MOLPRO 2010.1 \cite{molpro10_short}. The core electrons of both lithium and rubidium were replaced with the Stuttgart full-core relativistic (ECP2SDF and ECP36SDF) effective core potential \cite{fuentealba1982,szentpaly1982,fuentealba1983}. 
This reduces the calculation to a two electron problem while modeling the relativistic contribution to the rubidium core. To accurately describe the first excited states of LiRb, the optimized valence basis sets of Aymar and Dulieu \cite{Dulieu:JCP:2005,aymar2006} were used in all calculations. The effects of including additional diffuse functions were considered, but were found to not be large enough to be included in the calculation.
%end Jason Byrd

\subsection{Pump pulse and formation of a wave packet}

We proceed to simulate the two-photon ``pump-dump'' photoassociation process using a time-dependent wave packet approach (TDWP) \cite{SG:NJP:2009,PhysRevA.73.033408,2006JPhB...39.1017K}. The starting point in the TDWP are two ultracold atoms colliding with relative kinetic energy $E$ in an external magnetic field. As shown in the previous section, the collisional wave function is composed of several hyperfine channels and depends on the magnetic field (Figure \ref{fig:wfs_singlet}).
In accordance with common experimental realizations we assume a spin-polarized ultracold gas, where the entire population has been transferred to the lowest-energy hyperfine state $|11\rangle$. The state $|11\rangle$ is coupled by hyperfine and magnetic interactions to seven other states, one of mainly singlet and six of mainly triplet character (Figure \ref{fig:hyperfine}). 
At the near-resonant magnetic field, $B=1067$ G, these states are detuned in energy between 0.1 and 0.3 cm$^{-1}$ from the initial state, implying that a short photoassociation ``pump'' pulse will non-resonantly couple the channels to each other as well as to the target vibrational states. Nevertheless, the interchannel couplings are not strong, and it is sufficient to consider a single initial state on the time scales used in this work.

The Hamiltonian describing the optical transition between the ground X$^1\Sigma^+$ and 
excited B$^1 \Pi$ electronic state can be represented in the diabatic basis as \cite{SG:NJP:2009,koch2012coherent}
\begin{eqnarray}
\Op{H}' = \left( \begin{array}{cc} 
\Op{T} + V_{{\rm X}^1\Sigma^+}(R) & \bm{\mu}(R) \cdot \bm{\varepsilon}^{\ast}(t) \\
\bm{\mu}(R) \cdot \bm{\varepsilon}(t) & \Op{T} + V_{{\rm B}^1\Pi}(R) - \hbar\omega_L \end{array} \right) \,,
\label{hamil}
\end{eqnarray}
where $\Op{T}$ is the kinetic energy operator, $V_{{\rm X}^1\Sigma^+}(R)$ and $V_{{\rm B}^1\Pi}(R)$ are the electronic potentials, $\bm{\mu}(R)$ is the corresponding electronic dipole transition moment, and $\bm{\varepsilon}(t)$ is the laser field polarization vector.
The dipole and rotating-wave approximations are used.
We assume a transform-limited pulse
\begin{equation}
   \varepsilon(t) = \varepsilon_0 f(t) \cos(\omega_L t) \,,
\end{equation}
with a Gaussian profile $f(t) = \exp[- \alpha(t-t_c)^2]$ centered at the time $t_c$. Here, $\omega_L/(2\pi)$ is the central frequency of the pulse, chosen (during the optimization steps) to be resonant with a particular vibrational level of the state B$^1\Pi$, detuned from the Li(2$S$ $^2S_{1/2}$) + Rb(5P $^2P_{1/2}$) atomic asymptote by $\delta_L$. 
The full-width-at-half-maximum (FWHM) of the intensity profile of the pulse, $\tau_L = (2 \ln 2 / \alpha)^{1/2}$, is selected such that the pulse excites several vibrational levels of the B$^1\Pi$ state, forming a wave packet.
The wave packet dynamics is obtained from the time-dependent Schr\"odinger equation ($\hbar=1$)
\begin{equation}
 i \frac{\partial}{\partial t} \Xi(t) = \Op{H}'(t)\Xi(t) \,,
 \label{eq:tdse}
\end{equation}
which is solved by expanding the time evolution operator $\hat{U}(t)=\exp[-i\Op{H}'t]$ in Chebychev polynomials \cite{Kosloff:ARPC:1994}. The pump and dump pulse are treated separately and any coherent effects are neglected.

For the pump pulse, the initial (or ``Feshbach molecule'') state wave function, $\tilde{\Xi}(t=0,R)$ for the magnetic field $B$, is constructed by projecting the component $|11\rangle$ of the total Feshbach molecule state $|\psi\rangle$, given in Eq. (\ref{eq:psi_coupled}), to the short-range uncoupled molecular basis $|S \, m_S \, I \, m_I \rangle$:
\begin{equation}
 \Xi(t=0,R) = \sum_{I,m_I} g_{I,m_I}(R)|0 \, 0 \, I \, m_I \rangle \,,
\end{equation}
where the coefficients $g_{I,m_I}(R)$ are defined as
\begin{equation}
 g_{I,m_I}(R) = \frac{\varphi_{11}(R)}{R} \langle 0 \,0 \, I \, m_I | 1 \,1\rangle \,,
\end{equation}
and the state $|11\rangle$ is given by Eq. \ref{eq:chi}.
To match the normalization used in the TDWP calculation, the resulting initial wave function is box-renormalized to the Fourier grid according to \cite{2004EPJD...31..239L}
\begin{equation}
  \Xi(0,R) = \left( \frac{dE_n}{dn} \right)^{-1/2} \tilde{\Xi}(0,R) \;,
\end{equation}
where $\frac{dE_n}{dn} = \frac{dE}{dn} \lvert_{E=E_n} = E_{n+1} - E_n$ is the density of states calculated from the energy spacings between two neighboring energy levels of the discretized continuum.

The dynamics of the excited state population is subsequently analyzed by observing time-evolution of the wave packet projections 
\begin{equation}
 P_{\mathrm{B}^1\Pi}^{v^\prime}(R,t) = \vert \langle \phi_{B^1\Pi}^{v^\prime} (R) \vert \bm{\mu}(R) \vert \Xi(R, t) \rangle \vert^2 \,,
 \label{eq:Pexcited}
\end{equation}
where $\phi_{B^1\Pi}^{v^\prime}(R)$ is the wave function of the vibrational level $v'$ of the B$^1\Pi$ electronic state. 
The total projection is obtained as a sum over all vibrational levels, $P_{\mathrm{B}^1\Pi}(R,t) = \sum_{v'} P_{\mathrm{B}^1\Pi}^{v^\prime}(R,t)$.

The optimization of the pump pulse parameters, including the magnetic field $B$, is performed in several self-consistent steps. Guided by similar studies \cite{PhysRevA.73.033408,SG:NJP:2009}, we select realistic but arbitrary pulse duration and intensity, while keeping the frequency red-detuned from the dissociation threshold of the B$^1\Pi$ state, and calculate the population transfer for a number of magnetic fields in the vicinity of the Feshbach resonance at $B_0=1067.85$ G. 
We find that the magnetic field $B_{\mathrm{res}}=B_{0}-0.15$ G = 1067.7 G yields an optimal ``photoassociation window'' with respect to higher vibrational levels ($v'>30$) of the excited state B$^1\Pi$. This magnetic field corresponds to a loosely bound Feshbach molecule initial state, for which the scattering length is large and positive (Figure \ref{f:resonances}).

\begin{figure}[tb]
 \centering
 \includegraphics[clip,width=\linewidth]{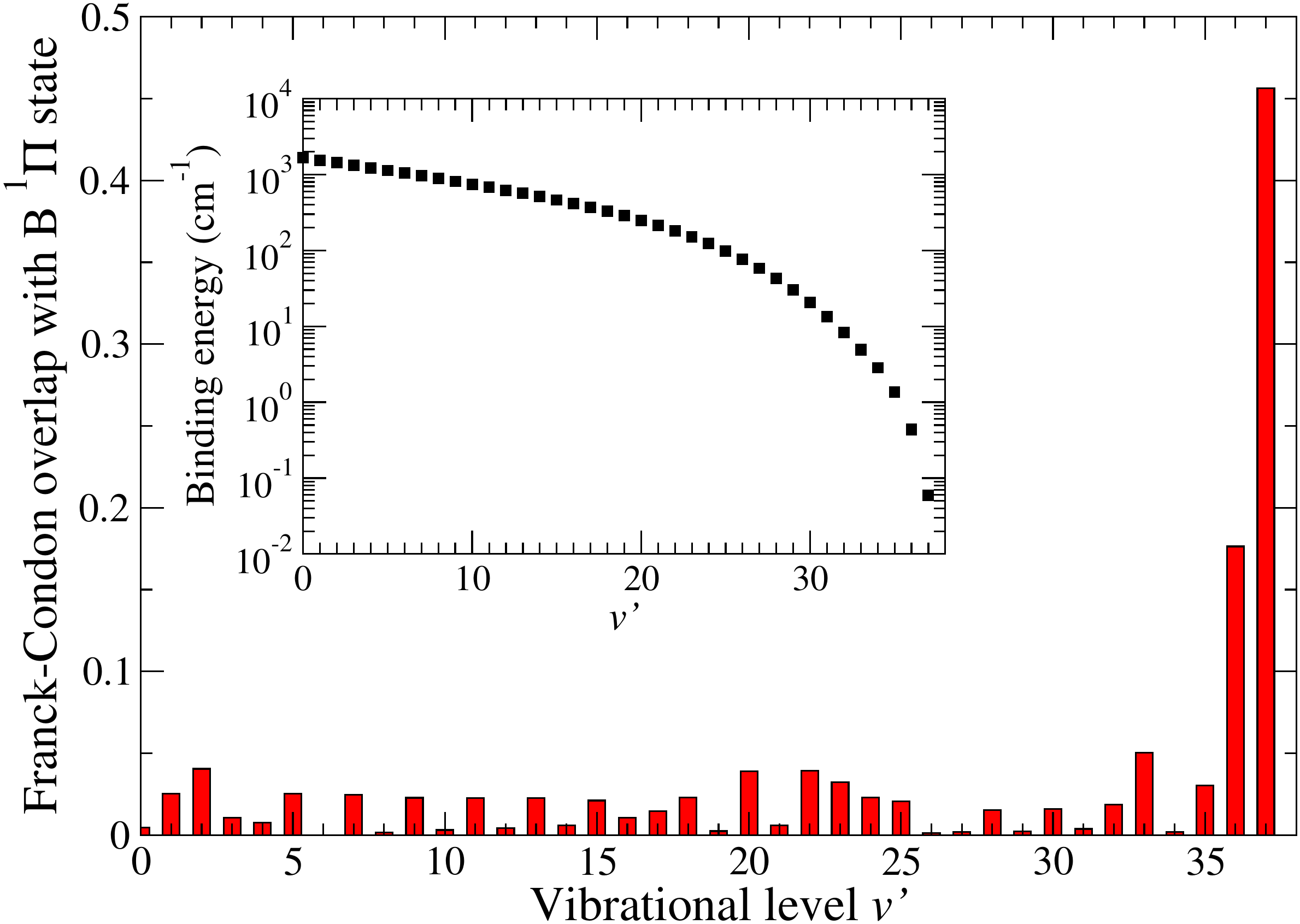}
 \caption{Franck-Condon overlap of the initial scattering wave function for $B=1067.7$ G with the vibrational levels of the B$^1\Pi$ state. \textit{Inset:} Binding energy of vibrational levels used in the calculation.}
 \label{fig:fcf1}
\end{figure}

Once the optimal magnetic field is determined, we proceed to optimize the pump pulse width, detuning, and duration of the so that the population transfer to the excited state is maximized.
To obtain general guidelines for the pulse parameters, we calculate the Franck-Condon factors between the initial state and bound vibrational levels of the B$^1\Pi$ state for $B=1067.7$ G (Figure \ref{fig:fcf1}). 
This is accomplished by diagonalizing the Hamiltonian in Eq. (\ref{hamil}) with the laser pulse turned off ($\varepsilon = 0$) and yields the binding energies and wave functions of the 50 bound vibrational levels of the ground X$^1\Sigma^+$ state and 38 bound vibrational levels of the excited B$^1\Pi$ state. The diagonalization is performed using mapped Fourier grid method \cite{Kokoouline:JCP:1999} with 511 points, extended up to $R \sim 13300$ Bohr. We find that so-called ``ghost levels'' are eliminated from the dynamics for the Sine basis representation. 

The Franck-Condon overlap is the best for the last two vibrational levels, $v'=36$ and $v'=37$. These levels, however, cannot be used to form the wave packet due to the fact that a large fraction of the pulse is being spent on the excitation of the continuum, resulting in dissociation of molecules \cite{2004EPJD...31..239L}. Thus, we select the central frequency of the pulse to be resonant with the level $v'=33$, detuned by $\delta$ = 4.5 cm$^{-1}$ from the Li(2$S$)+Rb(5$P$) asymptote.
Finally, a good choice of the temporal width of the pulse is $\tau_L=10.00$ ps. This corresponds to the spectral frequency width $\delta_\omega/(2 \pi c) = 2 \ln (2) / (\pi \tau_L c) = 1.47$ cm$^{-1}$, which is sufficiently broad to strongly excite vibrational levels $v'=33-35$ and produce a well-formed wave packet.
If we assume the illuminated volume to be within a sphere of 100 micron radius, the required energy per pulse is 20.18 nJ.
The pulse parameters are summarized in Table \ref{tab:pulse}.

\begin{figure}[t]
 \centering
 \includegraphics[clip,width=\linewidth]{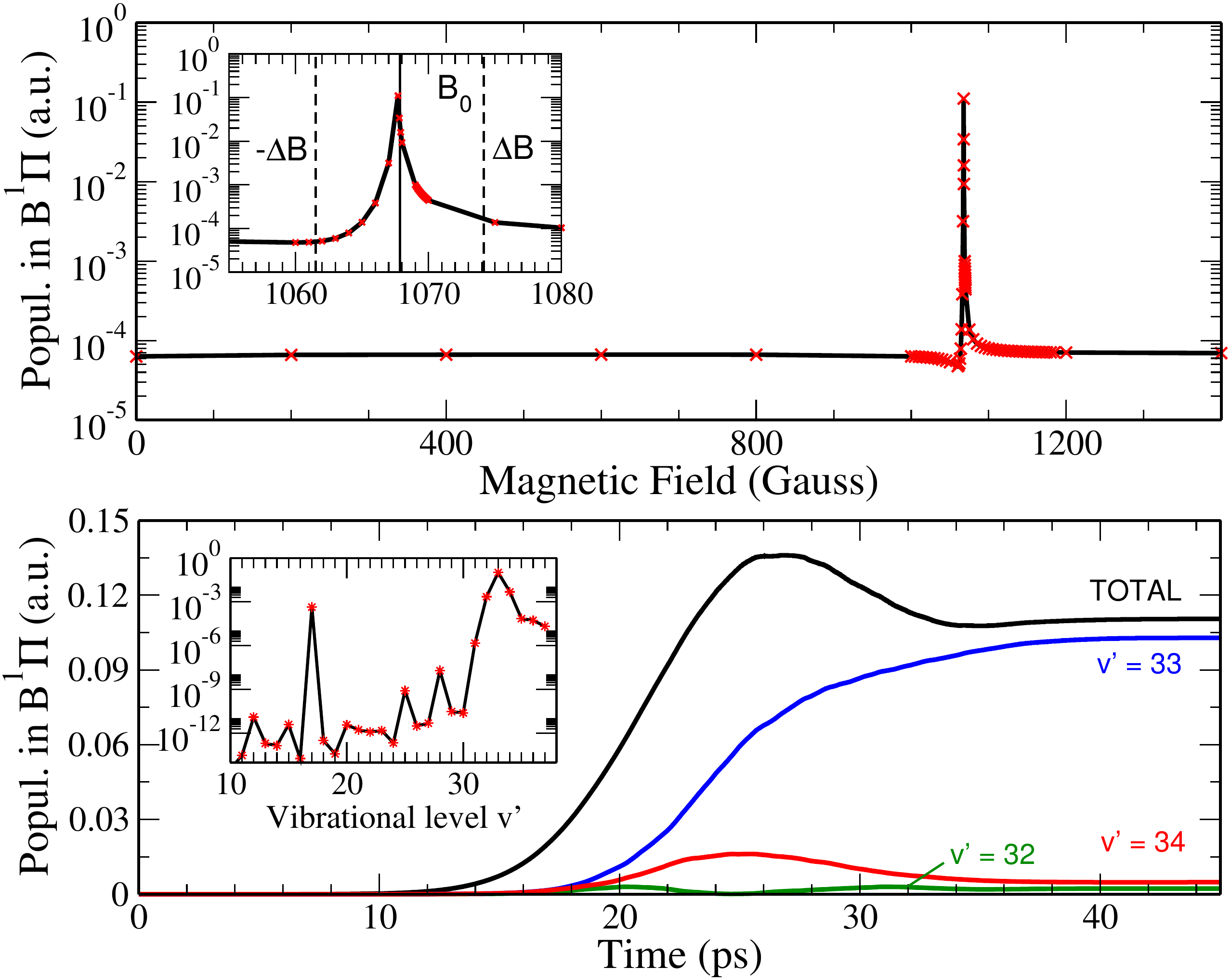}
 \caption{\textit{Top:} The final population in the B$^1\Pi$ electronic state after photoexcitation with the open channel singlet component of the initial wave functions calculated at different magnetic fields. \textit{Inset}: Zoom in on the resonance region. \textit{Bottom}: Time-dependence of the total population and contributions from vibrational levels $v^\prime$ = 32 - 34 during photoexcitation for the initial wave function at B = 1067.7 G. \textit{Inset}: population distribution of the B$^1\Pi$ state after the pulse.}
 \label{fig:pump-comp1}
\end{figure}

The fact that we neglect the other hyperfine states in our model warrants a more detailed explanation. Small energy separations between these states imply that they will be coupled by the laser pulse in the near-resonant regime. The resulting population cycling would be highly unfavorable as it would reduce the overall efficiency of the process and heat the ultracold gas cloud.
To justify the validity of the single-state approximation in the vicinity of the Feshbach resonance, we calculate the Rabi frequencies for the selected initial and final states, $v'=33$ of the B$^1 \Pi$ electronic state, $\Omega_R^{(11)} = \varepsilon |\langle 11 | \mu(R) | B^1 \Pi, v'\rangle|$, and $\Omega_R^{(17)} = \varepsilon |\langle 17 | \mu(R) | B^1 \Pi, v'\rangle|$ for the dipolar transition between the mainly-singlet closed channel $|17\rangle$, nearest in energy to the selected initial state $|11 \rangle$ (see Fig. \ref{fig:hyperfine}), and the target vibrational state considered. For the optimized pump pulse parameters and the maximal amplitude, the ratio of Rabi frequencies is $\Omega_R^{(17)}/\Omega_R^{(11)} = 0.088$, and the population of the excited state can be estimated to be smaller than $10^{-4}$ after the pulse. A similar analysis gives comparable or smaller values for the other channels. In addition, the dipole transition moments between the hyperfine states of mostly triplet symmetry will be very small, additionally suppressing the population transfer.

Once the pump pulse parameters are optimized, we recalculate the population transfer for all magnetic fields and follow its time-evolution in the B$^1\Pi$ state during the first 100 ps. The total population transferred to the excited B$^1\Pi$ electronic state by the pump pulse is shown in Figure \ref{fig:pump-comp1} for a range of magnetic fields. In the vicinity of the Feshbach resonance at $B_0$ it exhibits an increase of up to three orders of magnitude. This is a direct consequence of the enhancement of the initial scattering wave function amplitude in the ``photoassociation window'' caused by the coupling with closed hyperfine channels \cite{FOPA:PRL:2008}.
In Figure \ref{fig:pump-comp1} (bottom panel) we illustrate the evolution of the excited population during the first 50 ps for the optimal magnetic field $B_\mathrm{{res}}$ for significantly populated vibrational levels.
Note that while the levels $v'=32-34$ are transiently strongly excited, only a fraction of the population ($\sim$ 0.11), mainly in $v'=33$, remains in the excited state after the pulse.

\begin{figure}[t]
 \centering
 \includegraphics[clip,width=\linewidth]{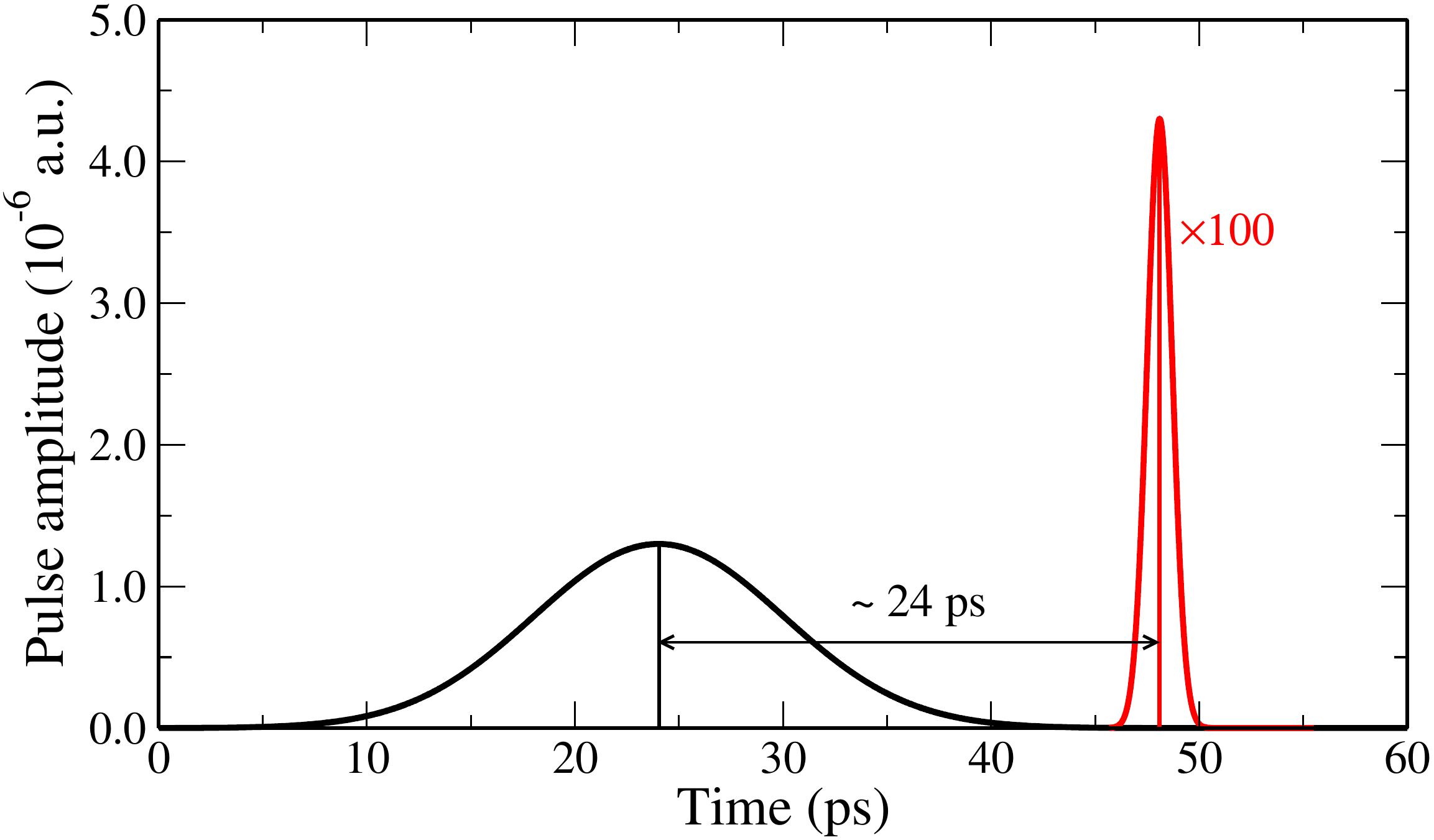}
 \caption{Laser pulse sequence after the optimization considered in the present study.}
 \label{fig:pulse}
\end{figure}

\subsection{Dump pulse and the ground state population}

\begin{figure}[t]
 \centering
 \includegraphics[clip,width=\linewidth]{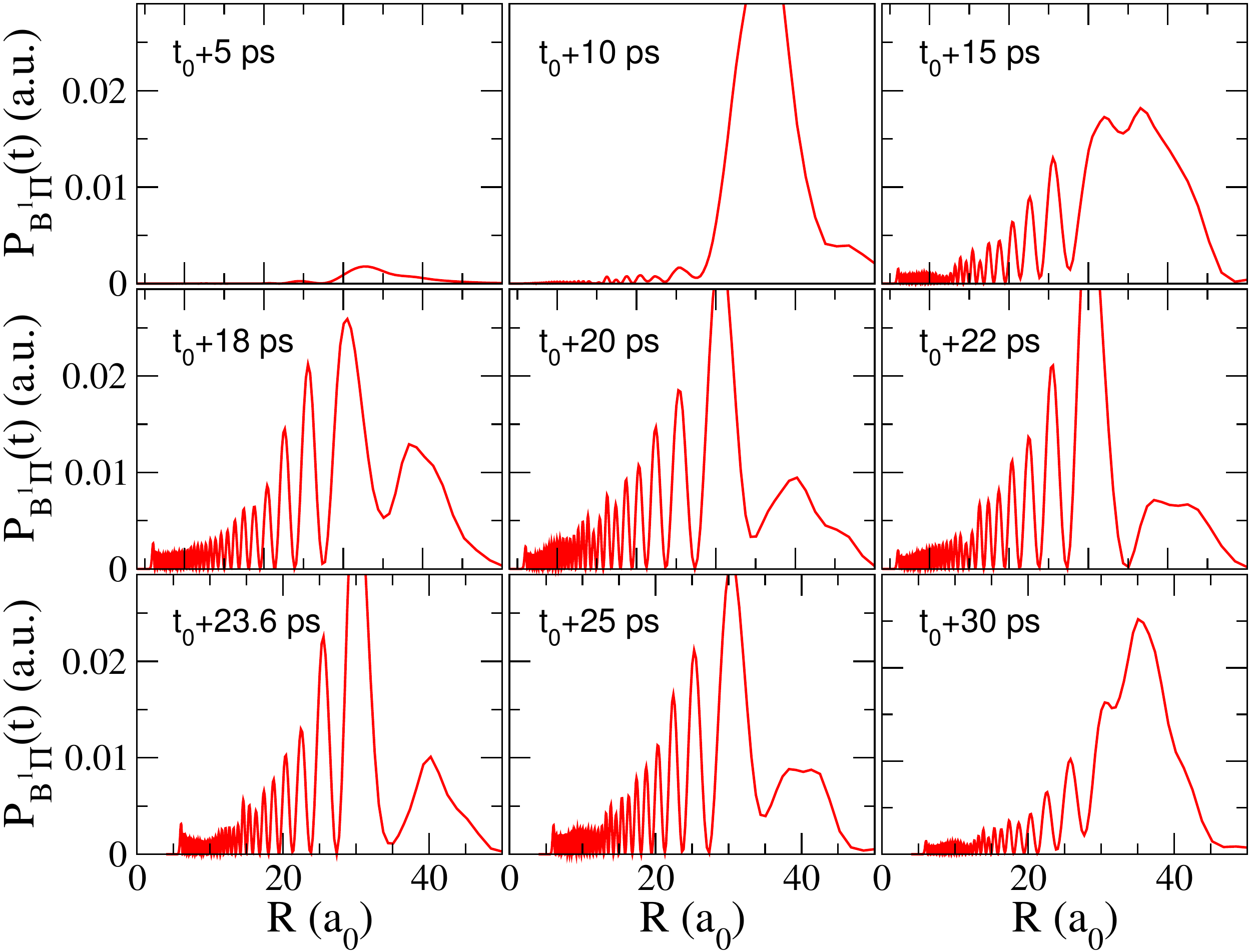}
 \caption{Snapshots of the total probability density distribution $P_{\mathrm{B}^1\Pi}$ of the wave packet from the time $(t_0+5)$ to $(t_0+30)$ ps, where $t_0=24$ ps corresponds to the pump pulse maximum.}
 \label{fig:wavepackets}
\end{figure}

The ``stabilization'' step of the two-photon photoassociation scheme, where the population is transferred back to the ground electronic state X$^1\Sigma^{+}$, is achieved by applying a second ``dump'' pulse after an optimal time delay (Figure \ref{fig:pulse}).
The optimized wave packet is allowed to propagate in the excited electronic state and the probability density distribution of the excited wave packet is monitored (Figure \ref{fig:wavepackets}).
Following the optimal short-range probability density of the excited wave packet is at 48 ps (Figure \ref{fig:wavepackets}, bottom middle panel), the dump pulse is initiated with a time delay of 24 ps from the maximum of the pump pulse sequence. Physically, this time delay corresponds to the time needed for the wave packet to reach the inner turning point. 

\begin{table}[b]
\centering
\begin{center}
\caption{Optimized pump and dump pulse parameters.
The detuning $\delta_L$ from the atomic asymptote, maximum electric field $\varepsilon_0$, temporal width (FWHM) $\tau_L$, spectral frequency $\delta_\omega$, and integrated pulse energy are given.}
\begin{tabular}{crrrrrr}
\hline \hline
 Pulse &  \multicolumn{1}{c}{$\delta_L$} & \multicolumn{1}{c}{$\varepsilon_0$} & \multicolumn{1}{c}{$\tau_L$} & \multicolumn{1}{c}{$\delta_\omega$/(2$\pi$c)} & \multicolumn{1}{c}{Eng/area}  \\
       &  \multicolumn{1}{c}{(cm$^{-1}$)} & (V m$^{-1}$) & (ps)     & (cm$^{-1}$)                 & (J m$^{-2}$ ) \\
\hline
 Pump  &    -4.5  & $1.34 \times 10^{6}$    & 10 &  1.47 &   0.05  \\
 Dump  &  4870.8  & $4.42 \times 10^{8}$    &  1 & 14.72 & 552.6   \\
\hline \hline
\end{tabular}
\label{tab:pulse}
\end{center}
\end{table}

The dump pulse parameters are also optimized to maximize the population transfer. Since the pulses do not overlap, the optimization can be performed separately and any coherent effects can be neglected.
The guidelines for optimizing the pulse duration and central frequency are the fact that the majority of the population in the excited state is comprised of $v'=33$, as well as the narrow time window during which the overlap is favorable. Conversely, an efficient dump pulse is required to be short and intense, while the detuning is adjusted to target a specific deeply bound vibrational level ($v'' = 0-20$) of the ground state.
The optimization yields the temporal width of the dump pulse to be $\tau_L = 1$ ps with maximum electric field amplitude $\varepsilon_0$ = $4.42 \times 10^8$ V m$^{-1}$. The optimized pulse parameters are given in Table \ref{tab:pulse}.

\begin{figure}[t]
 \centering
 \includegraphics[clip,width=\linewidth]{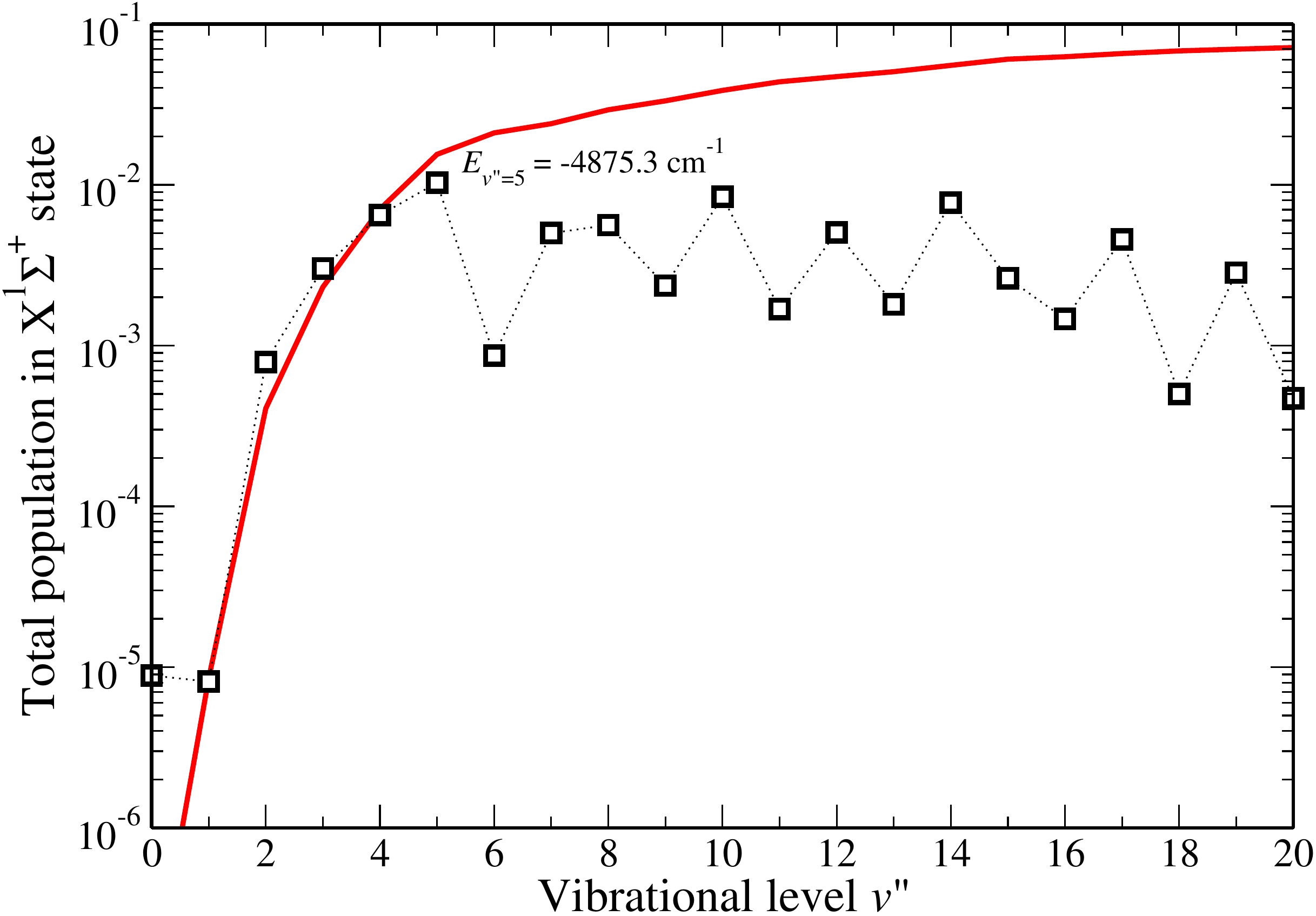}
\caption{Excited state population fraction transferred to a particular vibrational level $v^{\prime\prime}$ of the ground state by the dump pulse. Vibrational levels up to $v''=20$ and cumulative population starting from $v''=0$ (red line) are shown.}
\label{fig:dump}
\end{figure}

The resulting population in vibrational levels up to $v''=20$ of the ground state after the dump pulse is illustrated in Figure \ref{fig:dump}. 
Note that these are the final populations after the pump-dump pulse sequence.
For the optimized pulse parameters, the vibrational level $v''=5$, with the projection $P^{v''}_{X^{1}\Sigma}(t_f) \approx 0.011$, is most efficiently populated, followed by $v''=10$ and $v''=14$.
The deepest vibrational levels, $v''=0$ and $v''=1$, have particularly small projection integrals, $P^{v''=0}_{X^{1}\Sigma} \approx 8.9 \times 10^{-6}$ and $P^{v''=1}_{X^{1}\Sigma} \approx 8.1 \times 10^{-6}$. This is due to the unfavorable overlap between these levels and the inner turning point of $v'=33$ of the B$^{1}\Pi$ state. Vibrational levels higher than $v''=20$ are less efficiently populated as the wave packet is localized near the inner turning point.

\subsection{Total number of produced molecules}

The total number of molecules produced in photoassociation with short pulses has been estimated by Luc-Koenig et al. \cite{2004EPJD...31..239L} and specifically addressed in a recent publication by Koch et al. \cite{2006JPhB...39.1017K}. Both authors use a discretized quasicontinuum represented on a nonequidistant Fourier grid, an approach well suited to our proposed scheme. The quasicontinuum approximation makes it possible to represent the thermal gas as an ensemble of wave packets and directly apply statistical mechanics to find its average.

Following \cite{2006JPhB...39.1017K}, we estimate $N_{\mathrm{exc}}$, the number of molecules formed per pulse per photoassociated pair in the excited state $B^{1}\Pi$, 
\begin{equation}
  N_{\mathrm{exc}} = \frac{1}{2} N_{\mathrm{Li}}N_{\mathrm{Rb}} \langle P_e \rangle \;, 
  \label{eq:Psum}
\end{equation}
where $N_{\mathrm{Li}}$ and $N_{\mathrm{Rb}}$ are the numbers of Li and Rb atoms in the gas, and $\langle P_{e} \rangle = \sum_{v'j'} \langle P_{e}^{v'j'} \rangle$ is the total excited state population after the photoassociation pulse, expressed as a sum of thermally-averaged partial populations $P_{e}^{v'j'}$ for ro-vibrational levels $v'j'$, given in Eq. (\ref{eq:Pexcited}).
By following the thermal averaging procedure outlined in Ref. \cite{2006JPhB...39.1017K} and assuming a deep ultracold regime, where Wigner threshold law is valid and partial waves higher than the $s$-wave can be neglected, we arrive to a simplified expression for the partial thermally-averaged photoassociation probability
\begin{equation}
  \langle P_e^{v'j'} \rangle_{T \approx 0} = \frac{4 \pi}{V} \frac{P_e^{v'j'}(t_f)}{2 \mu E} \;.
  \label{eq:Ppartial}
\end{equation}
Here, $V$ is the trap volume and $\mu$ is the reduced mass, and $E$ is the relative collision energy. Note that Eq. (\ref{eq:Ppartial}) is valid for the initial wave function renormalized to have amplitude equal to unity instead of more common energy-normalization prefactor, resulting in the partial population matrix element $P_e^{v'j'}(t_f)$ being expressed in Bohr \cite{2004EPJD...31..239L,2006JPhB...39.1017K}. The angular factor, arising from integrating out the spherical harmonics coupled with the angular part of the excited state wave functions $\phi_e^{v'j'}$ and the electric dipole moment $\bm{\mu}$, is approximated as unity.

To estimate the total number of molecules transferred to the excited state, we assume that the initial cloud of trapped ultracold gas is at the temperature of 1 $\mu$K, has the volume $V = 0.1$ mm$^3$, contains $N_{\mathrm{Li}} = 10^6$ Li atoms, and $N_{\mathrm{Rb}} = 10^{10}$ Rb atoms. These physical parameters are comparable to the two-species MOT conditions reported in recent experiments \cite{koch2012coherent,doi:10.1021/cr300215h}.

The total population $P_e$ in the excited state, calculated in the previous section for the Feshbach-resonant magnetic field $B_{\mathrm{res}} = 1067.7$ G and and off-resonant magnetic field, $B_{\mathrm{off}}$, was found to be $P_{e}(B_{\mathrm{res}}) = 0.11$ and $P_{e}(B_\mathrm{off}) = 6.3 \times 10^{-5}$, respectively. This yields the total numbers of photoassociated molecules per a pump pulse (Eq. (\ref{eq:Psum})) equal to $N_{\mathrm{exc}}(B_{\mathrm{res}}) = 0.02$ and $N_{\mathrm{exc}}(B_{\mathrm{off}}) = 10^{-5}$, resulting in an enhancement of the photoassociation rate at the Feshbach resonance by three orders of magnitude. 
This is consistent with previous studies of photoassociation near a Feshbach resonance \cite{FOPA:PRL:2008,Elena:NJP:2009,2009NJPh...11e5047P}.

The total photoassociation probability per pump pulse should be compared to the estimates for RbCs and Cs$_2$. Assuming cesium MOT density of $10^{11}$ cm$^{-3}$, Koch \textit{et al.} predicted that up to 10 Cs$_2$ molecules can be photoassociated per an optimized chirped pulse \cite{2006JPhB...39.1017K}. If our estimates are rescaled to match these MOT conditions, we predict the formation of 9.5 molecules per pulse off-resonance, and about $1.7 \times 10^4$ molecules per pulse at the resonance.
These estimates are comparable with the rate of about one molecule per pulse for the gas density $10^{10}$ cm$^{-3}$ predicted for RbCs molecules \cite{SG:NJP:2009}.

We estimate the total number of molecules produced in the ground state per a pump-dump pulse sequence by multiplying the number of photoassociated molecules in the excited state with the probabilities to populate individual vibrational levels of the ground state (Figure \ref{fig:dump}). The resulting number of deeply bound ultracold molecules in the $v''=0-20$ is about $1.2\times 10^{-3}$ and $6.8\times 10^{-7}$ on- and off-resonance, respectively. The level $v''=5$ is the most efficiently populated, with about $1.7 \times 10^{-4}$ molecules formed per a pulse sequence initiated for the optimal magnetic field $B_{\mathrm{res}}$.
By the level of approximations introduced, the expression given in Eqs. (\ref{eq:Psum},\ref{eq:Ppartial}), is comparable to the scaling law estimate by Koch \textit{et al.} (shown in Figs. 8-10 of Ref. \cite{2006JPhB...39.1017K} as black lines). Therefore, we believe that the estimated total number of atoms in the trap remains the main source of uncertainty of our predictions.

\section{Summary and Conclusions}
\label{sec:conclusion}

We analyzed the effects of an external magnetic field tuned near a broad Feshbach resonance on the two-photon pump-dump photoassociation with short pulses of deeply bound ultracold molecules performed in a thermal atomic ultracold gas in the Wigner regime. The study was conducted on $^6$Li$^{87}$Rb molecule formed from a two-component ultracold gas of Li and Rb atoms via the intermediate B$^1\Pi$ excited electronic state. For this molecule, our results suggest that the photoassociation rate could be increased by about three orders of magnitude for the optimal near-resonant magnetic field, as compared to the field-free or off-resonant scenario.
The enhancement is largely caused by our choice of the loosely bound ``Feshbach molecule'' as the initial state during the pump pulse. In this regime, at near-resonant magnetic fields, energetically open and closed hyperfine components of singlet and triplet symmetry become strongly mixed, leading to an enhancement of the probability amplitude in the short range. The total wave function is effectively able to tunnel further inward, improving its overlap with excited electronic states.

The use of short pulses further optimizes the process and offers better control over the final population distribution. The pump pulse needs to be sufficiently short to prevent Rabi cycling of the transferred population back to the initial state, as well as sufficiently broad in energy space to create a well-formed wave packet in the excited state. The dump pulse is executed when the wave packet slows down and starts to reflect off the repulsive inner wall of the electronic potential. By iteratively modifying the pulse parameters in the calculation, we were able to further optimize the transfer rate to the low vibrational levels of the ground electronic state.

We estimated that an optimized photoassociation scheme would yield approximately 0.02 molecules in the excited electronic state per a pump pulse for the experimental MOT densities \cite{koch2012coherent,doi:10.1021/cr300215h}, or about $1.7 \times 10^4$ molecules per pulse for the MOT densities thought to be possible \cite{2006JPhB...39.1017K,SG:NJP:2009}. We found that an optimized dump pulse that efficiently populates the lowest 20 vibrational levels of the electronic ground state results in about 7\% transfer efficiency, or production of about 100 ground state molecules per a pulse sequence, assuming more favorable MOT conditions.
To form a larger number of molecules in the ground state a pulse train needs to be applied. For example, $10^4$ pulse pairs with the temporal separation of 100 ps between the two pulse sequences would produce $10^6$ LiRb ground state molecules per 1 $\mu$s, providing all loss mechanisms were neglected. 
While this estimate is almost certainly overly optimistic, it does show that the proposed technique could produce a large ensemble of ultracold molecules. In fact, the enhancement of the photoassociation rate near the Feshbach resonance could prove sufficient to validate experimental realizations in the systems where STIRAP or other approaches are ineffective.

The proposed photoassociation technique could be further improved by the introduction of different pulse shapes. In particular, the effectiveness of chirped pulses in population transfer \cite{2001PhRvA..63a3412V,2004PhRvA..70c3414L,2004EPJD...31..239L,2006JPhB...39.1017K,PhysRevA.73.033408} has been demonstrated. 
It is reasonable to assume that a chirped dump pulse would be able to both increase the total population transfer to the electronic ground state and alter its distribution among the lowest vibrational levels. 
Alternatively, a pulse shaper could be used to optimize the pulses in a feedback loop while monitoring the number of created molecules in a desired state.

Finally, while this study was performed for LiRb molecule and a magnetic Feshbach resonance was used to engineer the initial state, the time-dependent pump-dump approach is applicable to other molecules, as long as a sufficiently broad Feshbach resonance can be found and accessed experimentally.

\begin{acknowledgments}
We thank I. Simbotin for fruitful discussions. The authors acknowledge support from the Department of Energy (RC, SG) and the National Science Foundation grant No. PHY-1101254 (MG). SG also acknowledges support from DST Fast Track and BITS Pilani research initiation grant.
\end{acknowledgments}

\bibliography{lirb}

\end{document}